\RequirePackage{fix-cm}
\documentclass[sn-mathphys]{sn-jnl}      
\usepackage{graphicx}
\newcommand{\aap}{{Astron. Astrophys.}}
\newcommand{\apj}{{Astrophys. J.}}

\newcommand{\apjs}{{Astrophys. J., Suppl.}}
\newcommand{\apjl}{{Astrophys. J., Lett.}}

\newcommand{\araa}{{Ann. Rev. of Astr. \& Astrophys.}}

\begin{document}

\title{Inhomogeneity in the Local ISM and its Relation to the Heliosphere}




\author*[1]{\fnm{Jeffrey} \sur{Linsky}}\email{jlinsky@jila.colorado.edu}

\author[2]{\fnm{Seth} \sur{Redfield}}\email{sredfield@wesleyan.edu}

\author[3]{\fnm{Diana} \sur{Ryder}}\email{diana.ryder@colorado.edu}

\author[4]{\fnm{Eberhard} \sur{Moebius}}\email{emoebius@unh.edu}

\affil*[1]{\orgdiv{JILA}, \orgname{University of Colorado and NIST}, \orgaddress{\street{}, \city{Boulder}, \postcode{80309-0440}, \state{CO}, \country{USA}}}

\affil[2]{\orgdiv{Astronomy Dept. and Van Vleck Observatory}, \orgname{Wesleyan University}, \orgaddress{\street{Street}, \city{Middletown}, \postcode{06459-0123}, \state{CT}, \country{USA}}}

\affil[3]{\orgdiv{Department of Astrophysical and Planetary Sciences}, \orgname{University of Colorado}, \orgaddress{\street{}, \city{Boulder}, \postcode{80309-0391}, \state{CO}, \country{USA}}}

\affil[4]{\orgdiv{Space Science Center and Department of Physics}, \orgname{University of New Hampshire}, \orgaddress{\street{8 College Road}, \city{Durham}, \postcode{03824}, \state{NH}, \country{USA}}}

%


\abstract{This paper reviews past research and new studies underway of the local interstellar environment and its changing influence on the heliosphere.  The size, shape, and physical properties of the heliosphere outside of the heliopause are determined by the surrounding environment - now the  outer region of the Local Interstellar Cloud (LIC). The temperature, turbulence, and velocity vector of neutral atoms and ions in the LIC and other partially ionized interstellar clouds are measured from high-resolution spectra of interstellar absorption lines observed with the STIS instrument on the {\em HST}. Analysis of such spectra led to a kinematic model with many interstellar clouds defined by velocity vectors derived from radial velocity measurements. This analysis identified fifteen clouds located within about 10~pc of the Sun and their mean temperatures, turbulence, and velocity vectors. With the increasing number of sight lines now being analyzed, we find that temperatures and turbulent velocities have spatial variations within the LIC and other nearby clouds much larger than measurement uncertainties, and that these spatial variations appear to be randomly distributed and can be fit by Gaussians. The inhomogeneous length scale is less than 4,000 AU, a distance that the heliosphere will traverse in less than 600 years. The temperatures and turbulent velocities do not show significant trends with stellar distance or angle from the LIC center. If/when the Sun  enters an inter-cloud medium, the physical properties of the future heliosphere will be very different from the present. For the heliosheath and the very local interstellar medium (VLISM) just outside of the heliopause, the total pressures are approximately equal to the gravitational pressure of overlying material in the Galaxy. The internal pressure in the LIC is far below that in the VLISM, but there is an uncertain ram pressure term produced by the flow of the LIC with respect to its environment.}

\keywords{Stellar-interstellar interactions(1576), Interstellar clouds(834), Interstellar medium wind(848), Heliosphere (711), Heliopause(707), Warm neutral medium (1789), Ultraviolet sources (1741)}

\maketitle

\section{Interactions of the Outer Heliosphere with Local ISM}

     The heliosphere does not live in static isolation, but is instead encapsulated by an environment with inhomogeneous spatial properties that manifest as time variable external properties as the heliosphere traverses the interstellar medium. Total pressure balance between the outer heliosphere, hereafter called the very local interstellar medium (VLISM),  and the surrounding interstellar medium can greatly alter the size and shape of the heliosphere. For example, Zank \& Frisch (1999) and M\"uller et al. (2006)  computed heliospheric models for a wide range of interstellar pressures showing that when the heliosphere enters a cold cloud with density 100,000 times larger than the Local Interstellar Cloud (LIC), the heliosphere would shrink to the size of the inner solar system. The density and ionization within the heliosphere will respond to time variations in the interstellar density, ionization, flow vector, and magnetic field strength. In order to estimate the possible range of physical parameters in the heliosphere over time, it is essential to explore the past and future interstellar environments that the heliosphere has and will encounter. Our understanding of heliospheric evolution provides the basis for modeling astrospheres and their interactions with exoplanet atmospheres. 
     
     This paper views the heliosphere from the perspective of its changing environment. We review and critique past work on the local interstellar medium (LISM), and include a preview of work in progress concerning the inhomogeneity of the LISM (Linsky et al. 2022, Paper A) and pressure balance between the VLISM and the LIC (Linsky \& Moebius 2022, Paper B). Previous reviews include Redfield (2006) and Frisch et al. (2011).

\section{Observational Techniques}

     Our knowledge of the local interstellar medium (LISM) is based primarily on high-resolution ultraviolet spectra of stars that include narrow absorption lines produced by interstellar gas in the line of sight to the more distant star. The most useful interstellar lines are transitions from the ground states of H~I (Lyman-$\alpha$ 1215.67~\AA), D~I (Lyman-$\alpha$ 1215.34~\AA), Mg~II (2796.35, 2803.53~\AA), Fe~II (2586.65, 2600.127~\AA), and other lines of O~I, C~II, and Si~II. We fit Voigt profiles to the narrow interstellar absorption lines observed against broad emission lines formed in stellar chromospheres. We individually fit both fine structure components of the H~I and D~I Lyman-$\alpha$ lines. Interstellar absorption lines of Ca~II and Na~I in the optical spectrum are generally too weak to be seen on short sight lines. For every sight line that we have analyzed, there is at least one interstellar absorption component, but there are often two or more velocity components indicating different parcels of interstellar gas moving at different radial velocities in the sight line to the star. The average number of components is about 1.5 per sight line.

     The measurement of temperatures and turbulent broadening for a given velocity component requires high-resolution spectra of at least one low mass atom, typically D~I Lyman-$\alpha$, and at least one high mass ion such as Mg~II and Fe~II. We analyze the interstellar D~I line rather than the very optically thick H~I line. Since thermal broadening is proportional to mass $m^{-0.5}$ but turbulent (non-thermal) broadening is independent of mass, analysis of low and high mass species together are needed to separate the two broadening components. The inclusion of intermediate mass species when available leads to more accurate temperatures and turbulent velocities (Redfield \& Linsky 2004a).

     In these studies, we have assumed that the component of line broadening that depends on mass is thermal, so that we are measuring the gas temperature. Studies of the heliosphere indicate that small scale non-thermal processes such as shocks and charge exchange processes produce pickup ions that create a plasma with both thermal and supra-thermal velocities. Supra-thermal velocities will be most effective in broadening lines of low mass atoms such as D~I ($m$=2) relative to high mass ions such as Mg~II ($m$=24), because low mass means broader line profiles. Thus what we have called temperature may in fact be a combination of thermal and supra-thermal velocities, and that the inhomogeneous "temperatures" that we measure may in part be due to differences in the supra-thermal broadening. 
     
     The availability of high-resolution spectra of interstellar absorption lines with the GHRS and STIS instruments on the {\em Hubble Space Telescope (HST)} made it possible to infer temperatures and turbulent velocities for an increasing number of stars. Early analyses were for the brighter stars: Capella (Linsky et al. 1993), Sirius~A (Lallement et al. 1994), Procyon (Linsky et al. 1995), $\epsilon$~CMa (Gry et al. 1995, Gry \& Jenkins (2001), $\alpha$~Cen (Linsky \& Wood 1996), and $\beta$~CMa (Jenkins et al. 2000). With the increasing number of high-resolution spectra obtained by {\em HST} guest observer programs and from the {\em HST} archive, Redfield \& Linsky (2004b) systematically studied 50 velocity components in the sight lines to 29 stars. The mean weighted temperature for this sample is $6,680\pm 1,490$~K and the mean weighted turbulent velocity is $2.24\pm 1.03$ km~s$^{-1}$. For the 19 sight lines traversing the LIC, Redfield \& Linsky (2008) found mean weighted properties $T=7,500\pm 1,300$~K and $v$(turb)$ = 1.62\pm 0.75$~km~s$^{-1}$. Since then Gry \& Jenkins (2017) analyzed two velocity components towards $\alpha$~Leo, and Zachary et al. (2018) and Edelman et al. (2019) measured 11 components in the sight lines to 5 stars. In the following sections, we also include new measurements bringing the total number of analyzed sight lines to 58. Analyses of these sight lines is described in Paper A. In the following sections, we inter-compare the derived interstellar parameters for a total of 82 velocity components of which 34 are through the LIC. The sight lines extend to nearby stars, most of which are within 30~pc.

\section{The Kinematic Model of the Local ISM}

     Crutcher (1982) called attention to the coherent motion of nearby interstellar gas primarily in the hemisphere opposite to the Galactic Center. Analysis of the radial velocities of interstellar gas in the direction of 40 stars showed that interstellar gas is flowing away from the Scorpio-Centaurus OB Association, and that this flow vector is consistent with the observed flow of interstellar hydrogen and helium into the heliosphere. The flow vector identified by Crutcher (1982) is now called the Local Interstellar Cloud (LIC). Subsequent observations of optical interstellar absorption lines led to the identification of a second cloud vector centered in the Galactic Center direction, which Lallement \& Bertin (1992) called the Galactic cloud but is now called the G cloud. Additional observations led to a more complex interstellar cloud environment (e.g., Lallement \& Bertin 1992), but our present understanding of the local ISM required observations of a large number of sight lines with the high spectral resolution and velocity precision of the HRS and STIS spectrographs  on the {\em HST}. The many allowed transitions from highly populated ground energy states of elements and ions that are abundant in the ISM require the ultraviolet spectra that can be obtained with these instruments.
      
      In their analysis of 270 interstellar velocity components in the sight lines to 157 stars, Redfield \& Linsky (2008) noticed that over significant regions of the sky, the radial velocities of many absorption components are consistent with velocity vectors and assigned a cloud name to 15 of these vectors. The criteria for inclusion of a velocity component as a cloud member were: (1) that the measured radial velocity must be within 2 km~s$^{-1}$ (the velocity accuracy of the STIS instrument) of the projection of the velocity vector along the line of sight, and (2) that the sight line must be in close proximity to other sight lines with the same vector. The purpose of the second criterion is to identify only contiguous areas on the sky rather than isolated patches. They required at least 4 sight lines sufficiently separated in the sky to identify a cloud vector. Since they used radial velocity data, they could only identify clouds with large angular sizes to obtain the three dimensional velocity vectors. 
     
\begin{figure}[h]
\centering 
\includegraphics[angle=90,scale=0.5]{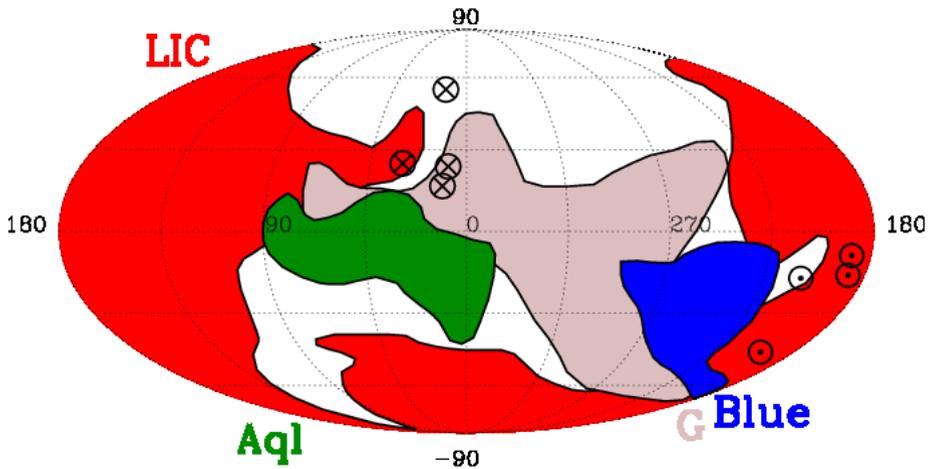}
\caption{Morphologies in Galactic coordinates of the four partially ionized LISM clouds 
  that are in contact with or close to the outer heliosphere. They are
  the LIC (red), which lies in front of $\epsilon$~Eri (3.2~pc), 
  the G cloud (brown), which lies in front of 
  $\alpha$~Cen (1.32~pc), the Blue cloud (dark blue), which lies in
  front of Sirius (2.64~pc), and the Aql cloud (green), which lies in front 
  of 61~Cyg (3.5~pc). The plot is in Galactic
  coordinates with the Galactic Center direction at the center. The
  upwind direction of the LIC velocity vector is
  indicated by the circled-cross symbol near $l=15^{\circ}$ and 
  $b=+20^{\circ}$, and the upwind directions of the other clouds have
  similar marks. The downwind
  directions are indicated by the circled-dot symbols. A full map
  of all 15 LISM clouds is given by Redfield \& Linsky (2008).
  Figure from Linsky et al. (2019).\label{allclouds}}
\end{figure}

     The 15 interstellar clouds identified by Redfield \& Linsky (2008) must be located within about 10~pc of the Sun because the interstellar absorption is in front of a nearby star at the end of the sight lines. A second reason is that Malamut et al. (2014) found that the number of interstellar absorption components does not increase with distance to the background star beyond 10~pc even for stars located as far as 60~pc. This group of interstellar clouds is now called the Cluster of Local Interstellar Clouds (CLIC), but there are undoubtedly many more clouds that could not be identified because of the small number of observed sight lines or small cloud angular extent due to small size or large distance. The many sight lines with interstellar absorption components not identified with known clouds indicates that there are many not yet identified clouds.        

      Figure~1 shows the locations and shapes of the four clouds closest to the heliosphere. The heliosphere is located just inside of the Local Interstellar Cloud (LIC) on the basis that (1) absorption produced by the LIC velocity vector is seen in slightly less than half of the sky (Redfield \& Linsky 2000, Linsky et al. 2019), and (2) that the speed and direction of gas flowing into the heliosphere as measured by the {\em Ulysses, STEREO}, and {\em IBEX} spacecraft is nearly the same as the mean flow parameters of the LIC. The small difference in speed and direction relative to the measured inflow velocity vector may indicate that the outer region of the LIC where the heliosphere now resides has slightly different properties than the mean values for the LIC (Linsky et al. 2019). Figure~2 shows the egg shape of the LIC in a plane parallel to the Galactic plane and contours in planes located above and below. 
      
\subsection{In its trajectory through the LISM, the future heliosphere depends on the future neutral hydrogen density}

     Figure~\ref{Suntrajectory} shows the trajectory of the heliosphere through nearby partially ionized interstellar clouds from 150,000 years ago until 50,000 years in the future (Vannier et al. in prep.).  The directions to the clouds  and the cloud velocities relative to the Sun are shown in the figure. The heliosphere entered the LIC roughly 60,000 years ago and will likely leave within the next 2,000 years. The extensions of the clouds along the sight lines are measured from the ratios of measured neutral hydrogen column densities to hydrogen number densities, $N$(H~I)/$n$(H~I), assuming that $n$(H~I)=0.20~cm$^{-3}$, the value at the edge of the LIC near the heliosphere (Slavin \& Frisch 2008). 
      
      The next closest cloud to the Sun is the G cloud observed in front of the nearest star system  $\alpha$~Cen (1.3~pc). Since interstellar lines seen in the $\alpha$~Cen sight line are centered at the G cloud velocity with no measured component at the projected LIC velocity, the heliosphere must be at the edge of the LIC. An upper limit to the hydrogen column density at the LIC velocity and the trajectory of the heliosphere roughly in the direction of $\alpha$~Cen, leads to the prediction that the heliosphere will leave the LIC in less than 2,000 years and then enter either the G cloud or an inter-cloud region that must be completely ionized so as to not be detected by interstellar Lyman-$\alpha$ absorption. In the former case the heliosphere will not greatly change, but in the latter case the changes will be profound as described in Section 6.

\begin{figure}
\includegraphics[angle=0,width=4.5in]{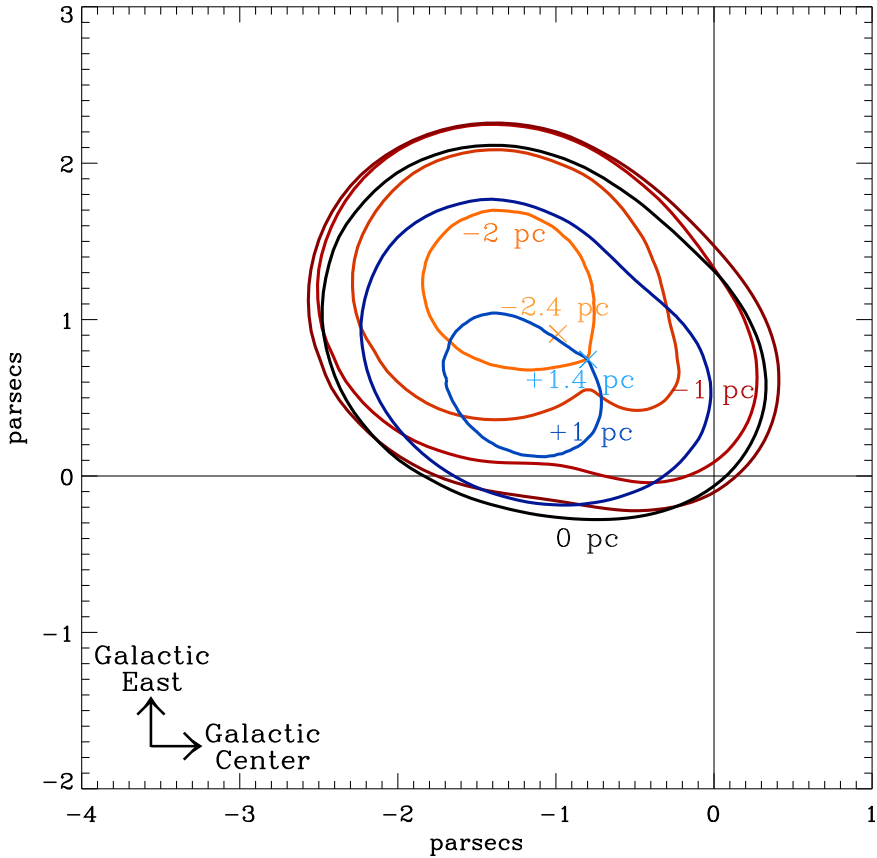}
\caption{Contour map of LIC as viewed from North Galactic Pole. The
  Sun is located at the origin (0,0). Red contours are cuts above the
  LIC center parallel to the Galactic plane, and the blue contours
  are cuts below. The X symbols indicate the locations where the
    edge of the LIC is furthest above or below the plane of the figure.
    Figure from Linsky et al. (2019).\label{fig:licconngp}}
\end{figure}

     The environment that the heliosphere will then encounter depends on the neutral hydrogen density in the G cloud. If $n$(H~I)$\approx 0.10$~cm$^{-3}$ in the G cloud, then the G cloud would be in direct contact with the LIC and the heliosphere will directly enter the G cloud, but for larger $n$(H~I) the heliosphere will enter an inter-cloud region before entering the G cloud. Thus only a factor of 2 uncertainty in the value of $n$(H~I) in the G  cloud will determine whether the future heliosphere will be similar to the present or become very different in less than 2,000~years. 
     
     There is no present way to measure $n$(H~I) in other clouds or indeed elsewhere within the LIC. If the assumption that $n$(H~I)$=0.20$~cm$^{-3}$ is valid for other clouds in the CLIC, then the trajectory of the heliosphere has been and will be through clouds with significant neutral hydrogen densities and through inter-cloud regions where the hydrogen is fully ionized. As outlined in Section 6, the properties of the heliosphere would be greatly different from the present in the absence of neutral hydrogen inflows from the LISM. If, instead, typical values of $n$(H~I) are~0.10 cm$^{-3}$ rather than 0.20~cm$^{-3}$, then the clouds would be a factor of 2 larger and completely fill the CLIC with no inter-cloud region of fully ionized hydrogen. If, on the other hand, $n(H~I)$ in nearby clouds is typically larger than 0.20~cm$^{-3}$, then a larger fraction of space would be devoid of neutral hydrogen and the heliosphere will spend a larger fraction of time in the inter-cloud medium.  
     
\begin{figure}[h]
\centering 
\includegraphics[width=11cm]{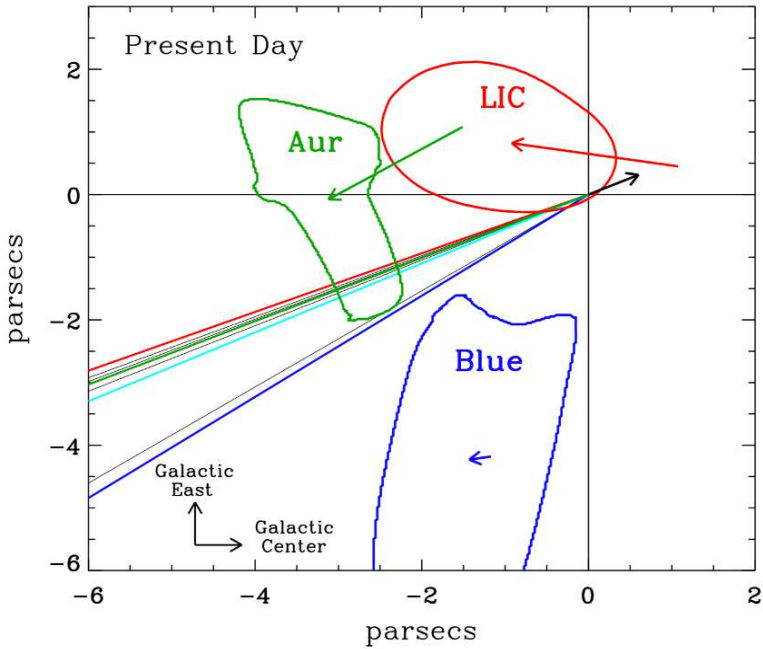}
\caption{The trajectory of the Sun through the local interstellar medium in the local standard of rest (LSR) from 150,000 years in the past to 50,000 years in the future. During this time the Sun passed through the Aur cloud, the Local Interstellar Cloud (LIC), and the intercloud medium. The LIC and two other clouds are shown with their velocities relative to the LSR indicated by colored arrows. The Sun's present position and direction are indicated by the black arrow. In the reverse directions are sight lines to stars that the Sun has passed. The heliosphere entered the LIC about 60,000 years ago and will leave within the next 2,000 years. The cloud morphologies and motions relative to the LSR are derived from their kinematics and reconstructions from many sight lines. The mp4 file for this movie is publicly available at https://jilafile.colorado.edu/index.php/s/wwUtbGsMUM22Vz2. Figure from Vannier et al. (in prep.).}\label{Suntrajectory}
\end{figure}

\subsection{How realistic is the multi-cloud model of the LISM?}

The multi-cloud model proposed by Redfield \& Linsky (2008) makes the simplifying assumption that each cloud moves rigidly though space with its own velocity vector and that each cloud has its own temperature, turbulent velocity, metal depletions, and boundaries that separate it from other clouds and the postulated inter-cloud medium. Gry \& Jenkins (2014) proposed a very different model in which the LISM consists of a single continuous cloud surrounding the Sun with an internal nonrigid flow and a gradient in metal depletion properties. They showed that this simpler one-cloud model fits the same data set used by Redfield \& Linsky (2008). 

Redfield \& Linsky (2015) then tested the ability of both models to fit a new data set of stars selected at random. This new data set was provided by an {\em HST} SNAP observing program in which high-resolution stellar spectra including the Mg~II, Fe~II, and Mn~II lines for a large number of stars were requested but the actual observations were selected according to the availability of single orbits between larger observing programs. The result was a random distribution of stars in Galactic coordinates. The SNAP data set consists of 76 velocity components in the sight lines to 34 stars (Malamut et al. 2014). Redfield \& Linsky (2015) found that the multi-cloud model provides a better fit to both the SNAP data set and the older data set. While the better fit of the available data by the multi-cloud model supports its continual use, future analyses of larger data sets may require a new morphology of the LISM with properties incorporating aspects of both models. 
     
\section{Inhomogeneity in Local Interstellar Clouds}

\subsection{Evidence for Inhomogeneity}

     The larger number of analyzed components now available enables a new study of the distribution of properties inside of the LIC and for the total data set including all clouds. This section is a progress report based on the number of sight lines analyzed to date, but more sight lines are being analyzed and the results will be published in Paper A. All of the included data have temperatures at least two or three times the measurement errors ($\sigma=2$ or $\sigma=3$ as indicated in the figures). Of the 89 velocity components measured so far, 58 meet this $\sigma=3$ requirement. For velocity components that traverse the LIC, we relax the selection criterion to $\sigma=2$ to increase the number of velocity components in the sample. It turns out that there is a wide distribution of both temperatures and turbulent velocities. 
     
\begin{figure}[h]
\centering
\includegraphics[width=10.0cm]{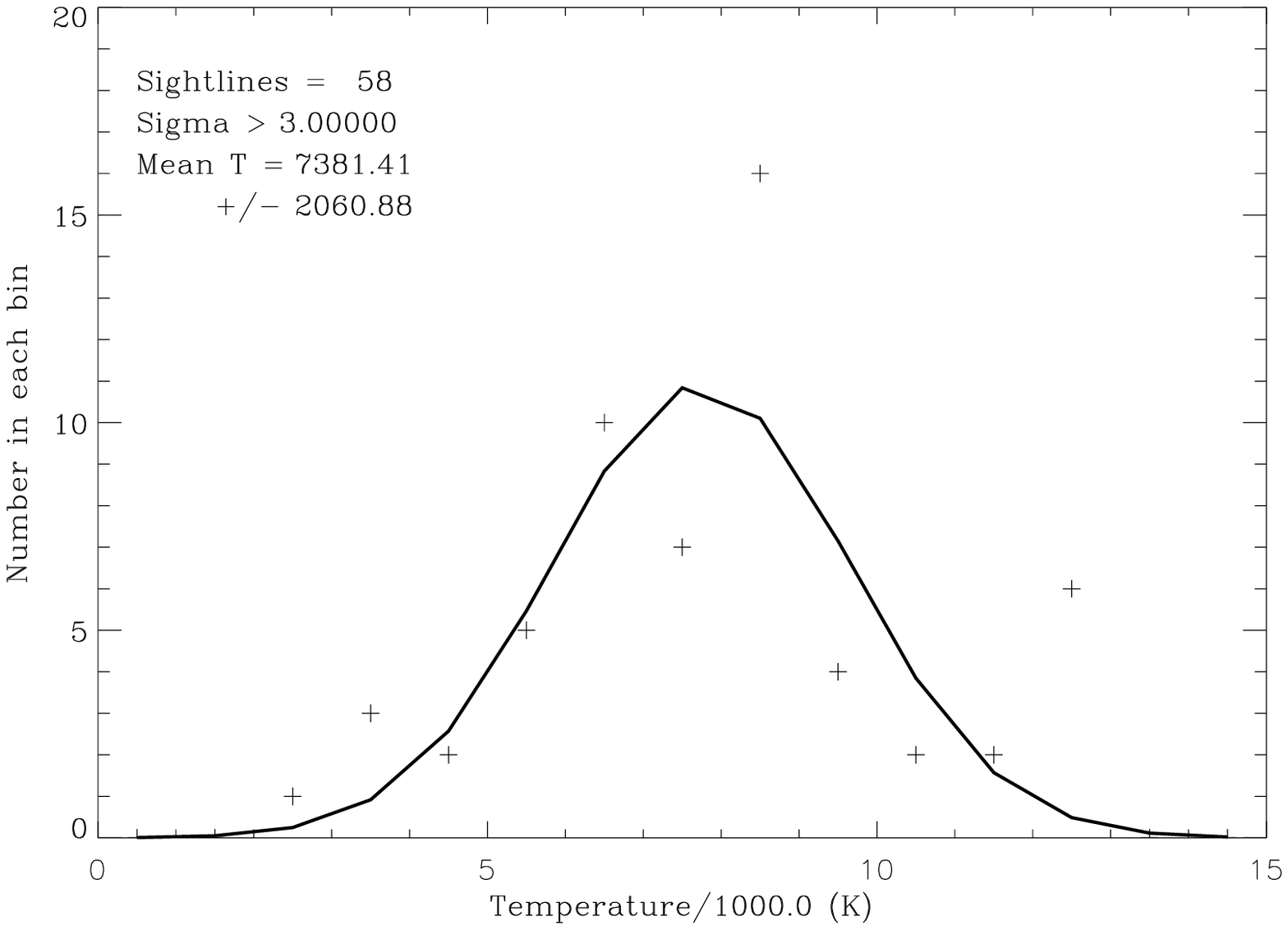}. 
\includegraphics[width=10.0cm]{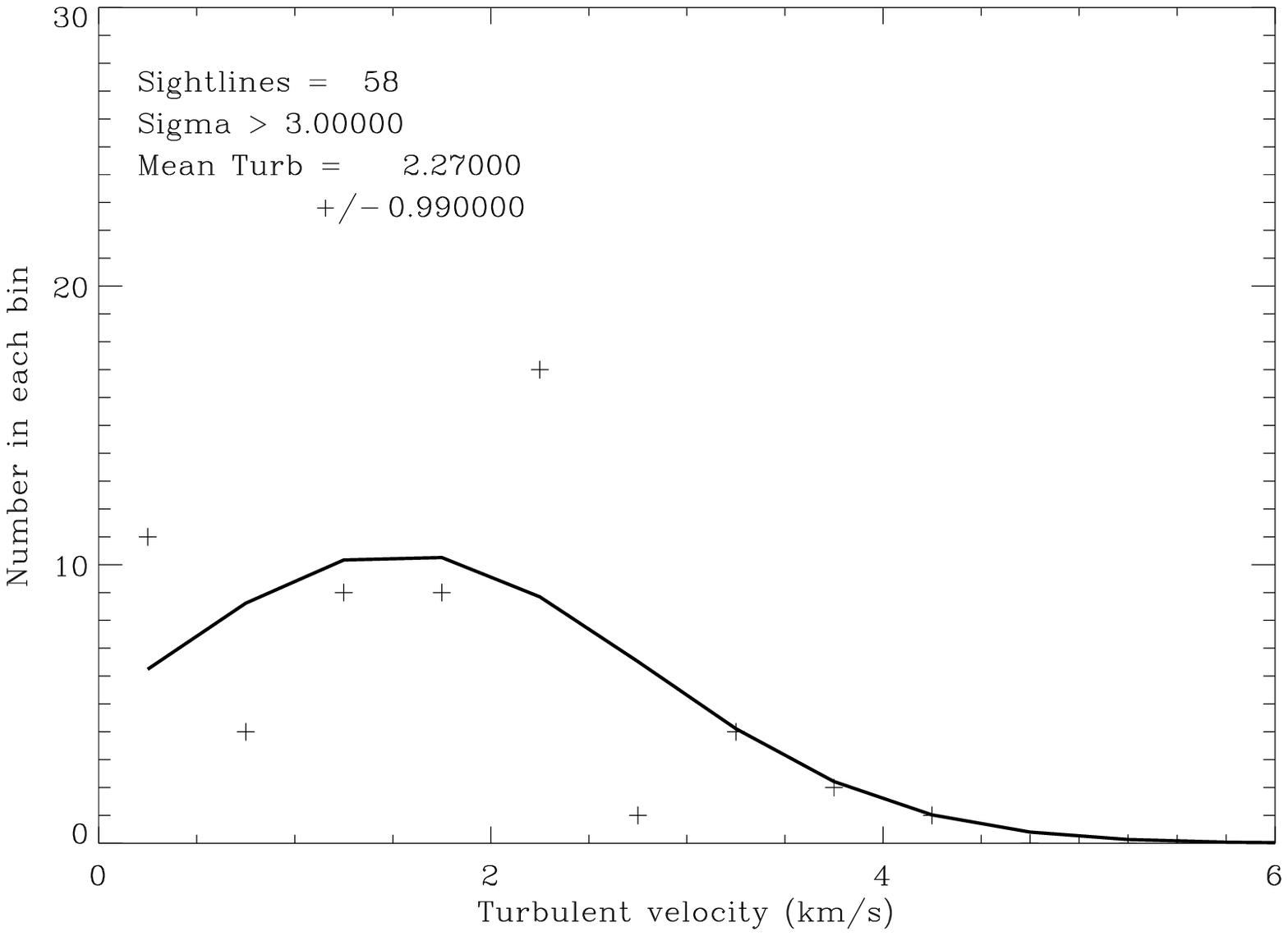}
\caption{Distribution of sight line temperatures ({\bf top}) and turbulent velocities ({\bf bottom}). Temperatures are binned in 1000~K intervals and turbulent velocities are binned in 0.5~km~s$^{-1}$ intervals. The solid curves are Gaussian fits to the weighted data.}\label{Tdist}
\end{figure}

     Figure~\ref{Tdist} shows the number of sight lines with temperatures within 1000~K wide bins. Although the mean temperature weighted by measurement uncertainties is $7381\pm2061$~K, the range of temperatures meeting the $\sigma=3$ criterion is from $2,450\pm700$~K for the LIC seen towards the star $\pi^1$~UMa (HD 72905) to $12,600\pm 2,400$~K for the Aql cloud seen towards $\alpha$~Aql (HD~187642). Figure~\ref{Tdist} shows that the temperatures appear to be randomly distributed with a range far in excess of measurement uncertainties. The temperatures can be fit by a Gaussian distribution, although there may be a significant number of outliers in the highest temperature bin 12,000--13,000~K. The distribution of turbulent velocities also covers a wide range about a mean value of $2.27\pm0.99$~km~s$^{-1}$ and is also fit by a Gaussian distribution. 

\begin{figure}[h]
\centering
\includegraphics[width=9.8cm]{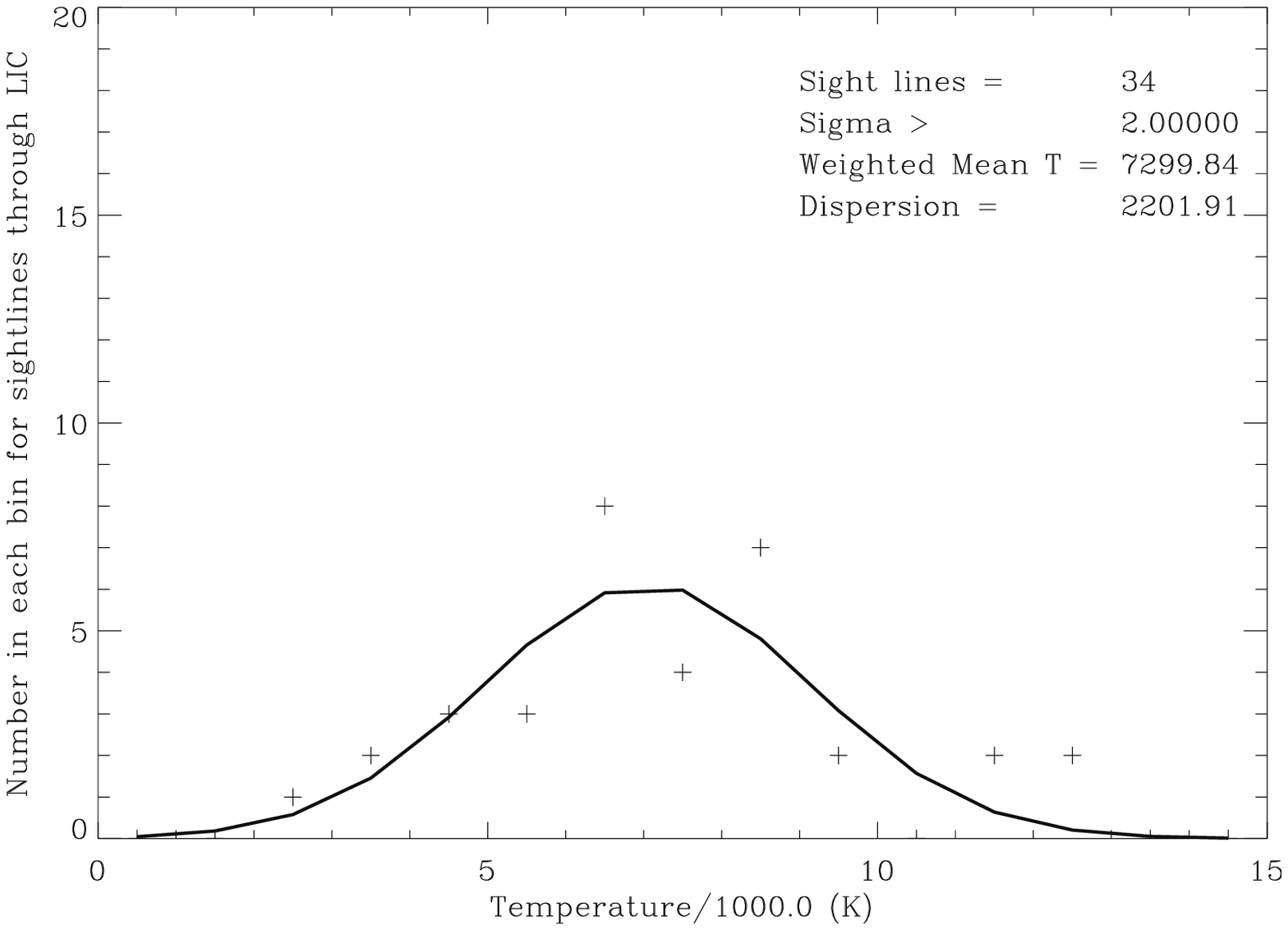}
\includegraphics[width=10.0cm]{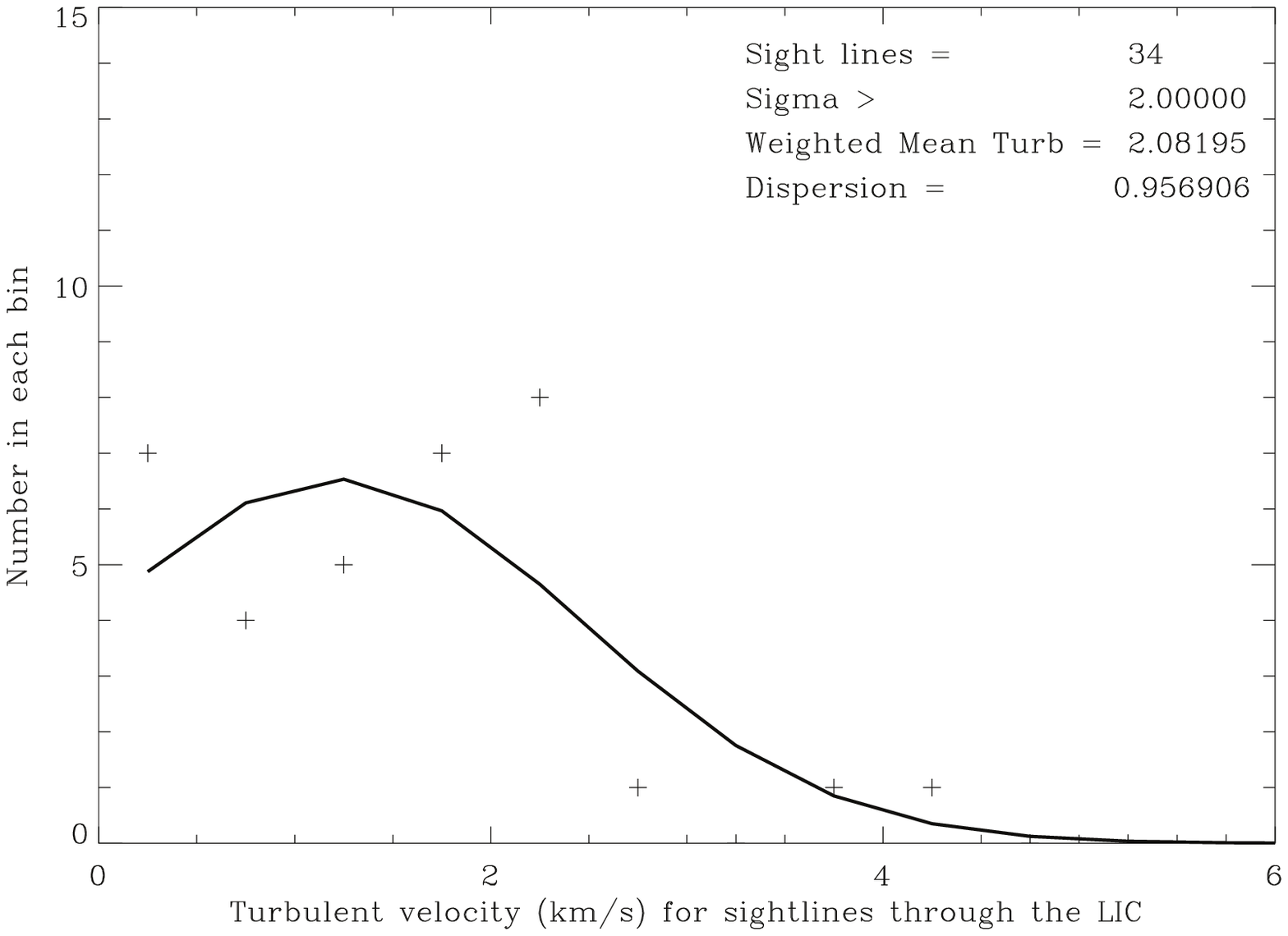}
\caption{Distribution of LIC sight line temperatures ({\bf top}) and turbulent velocities ({\bf bottom}). Temperatures are binned in 1000~K intervals and turbulent velocities are binned in 
0.5 ~km~s$^{-1}$ intervals.The solid curves are Gaussian fits to the weighted data.}\label{TLICdist}
\end{figure}

Figure~\ref{TLICdist} shows that the temperatures and turbulent velocities for the 34 LIC sight lines ($\sigma=2$) show distributions similar to the entire data set. The temperatures are also fit by a Gaussian distribution. The mean temperature weighted by the measurement uncertainties is $7,300\pm2202$~K and the mean weighted turbulent velocity is 
$2.08\pm0.96$~km~s$^{-1}$. Thus the LIC has a range of physical properties similar to the other clouds in the CLIC. 

\subsection{Are there systematic trends in temperature and turbulent velocity with spatial location?}

\begin{figure}[h]
\centering
\includegraphics[width=10.0cm]{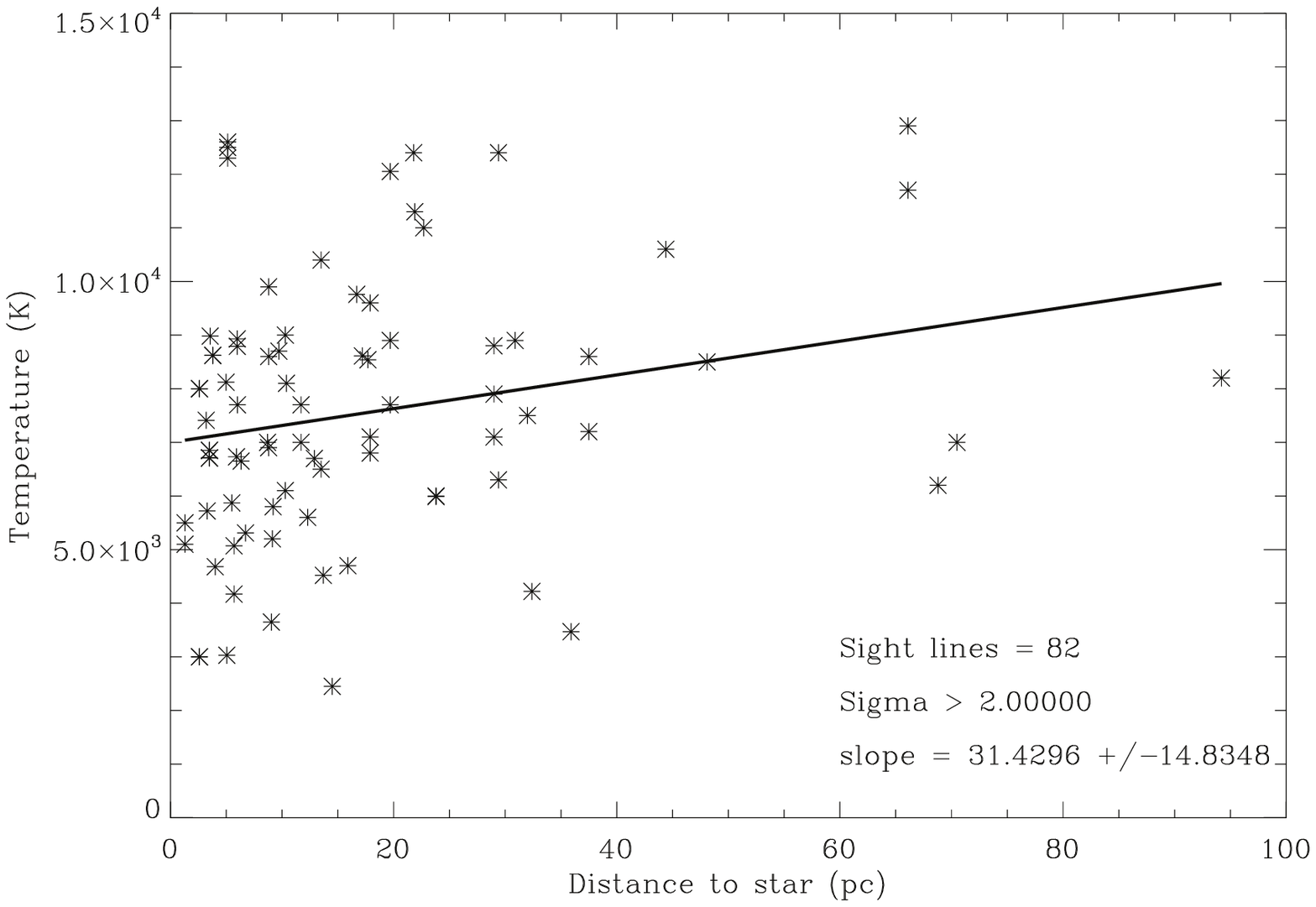}
\includegraphics[width=9.6cm]{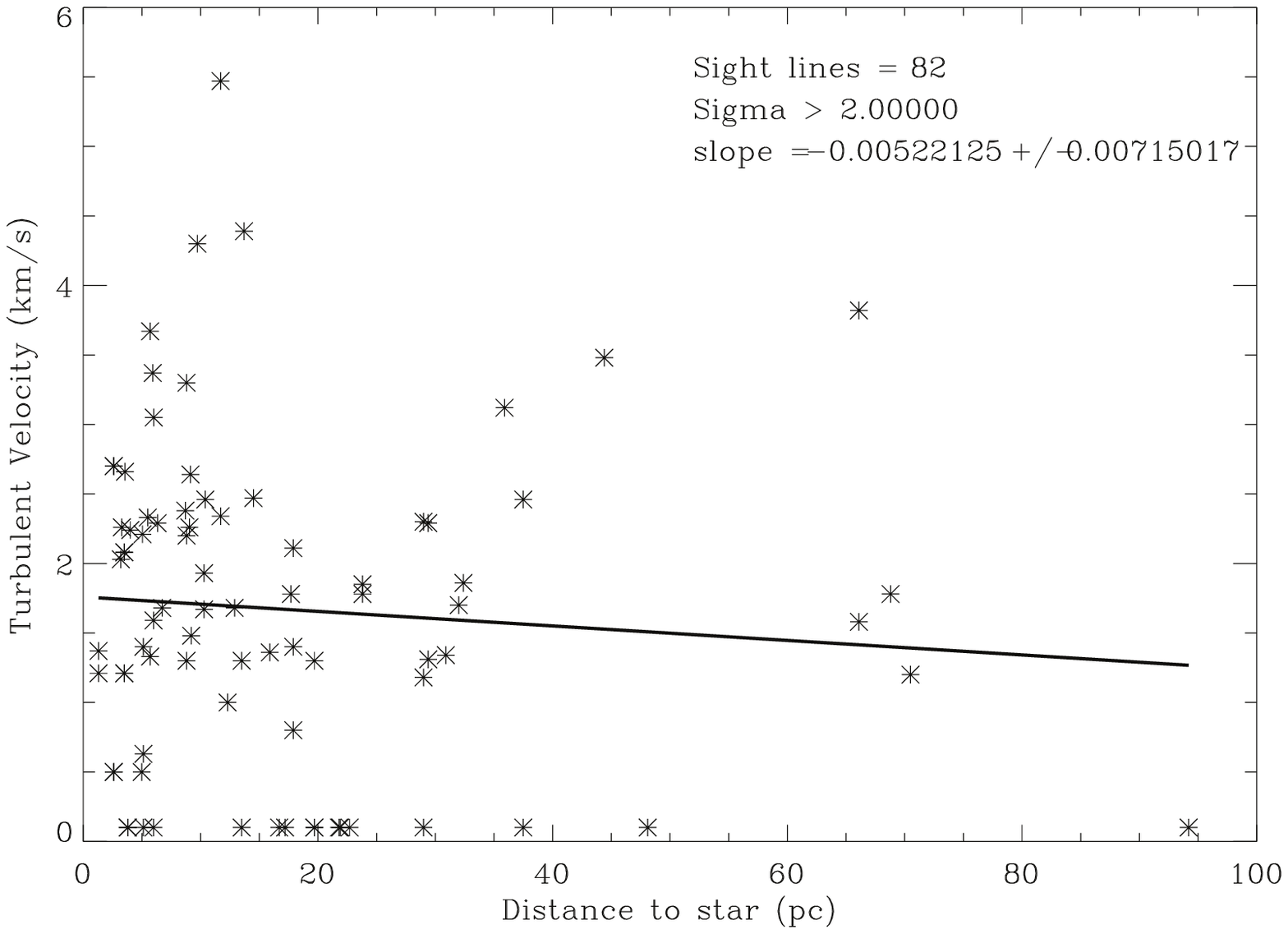}
\caption{{\bf Top}: Plot of sight line temperatures vs. distance to the star. The solid line linear fit to the data is $T=(7,460\pm 433) + (26.8\pm 16.8)d$ and thus not significant for demonstrating an increase or decrease with distance. {\bf Bottom}: Plot of sight line turbulent velocities vs. distance to the star.}
\label{Tvsd}
\end{figure}

We first consider whether temperatures and turbulent velocities tend to increase or decrease with distance to the star at the end of the sight line. Figure~\ref{Tvsd} shows no trend of temperatures or turbulent velocities with distance. The absence of trends indicates that the background star does not usually influence the foreground interstellar clouds in its sight line.

\begin{figure}[h]
\centering
\includegraphics[width=10.5cm]{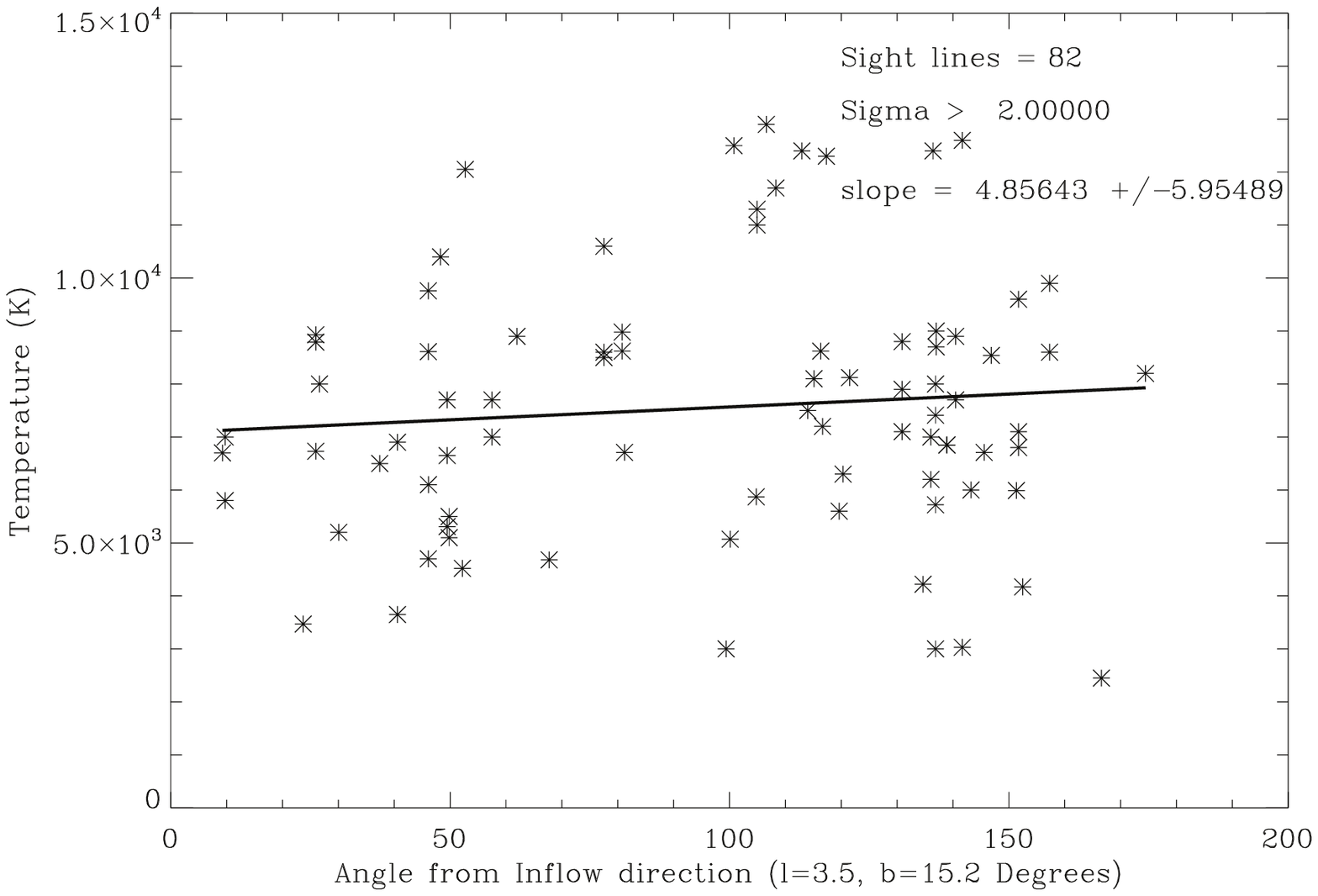}
\includegraphics[width=10.0cm]{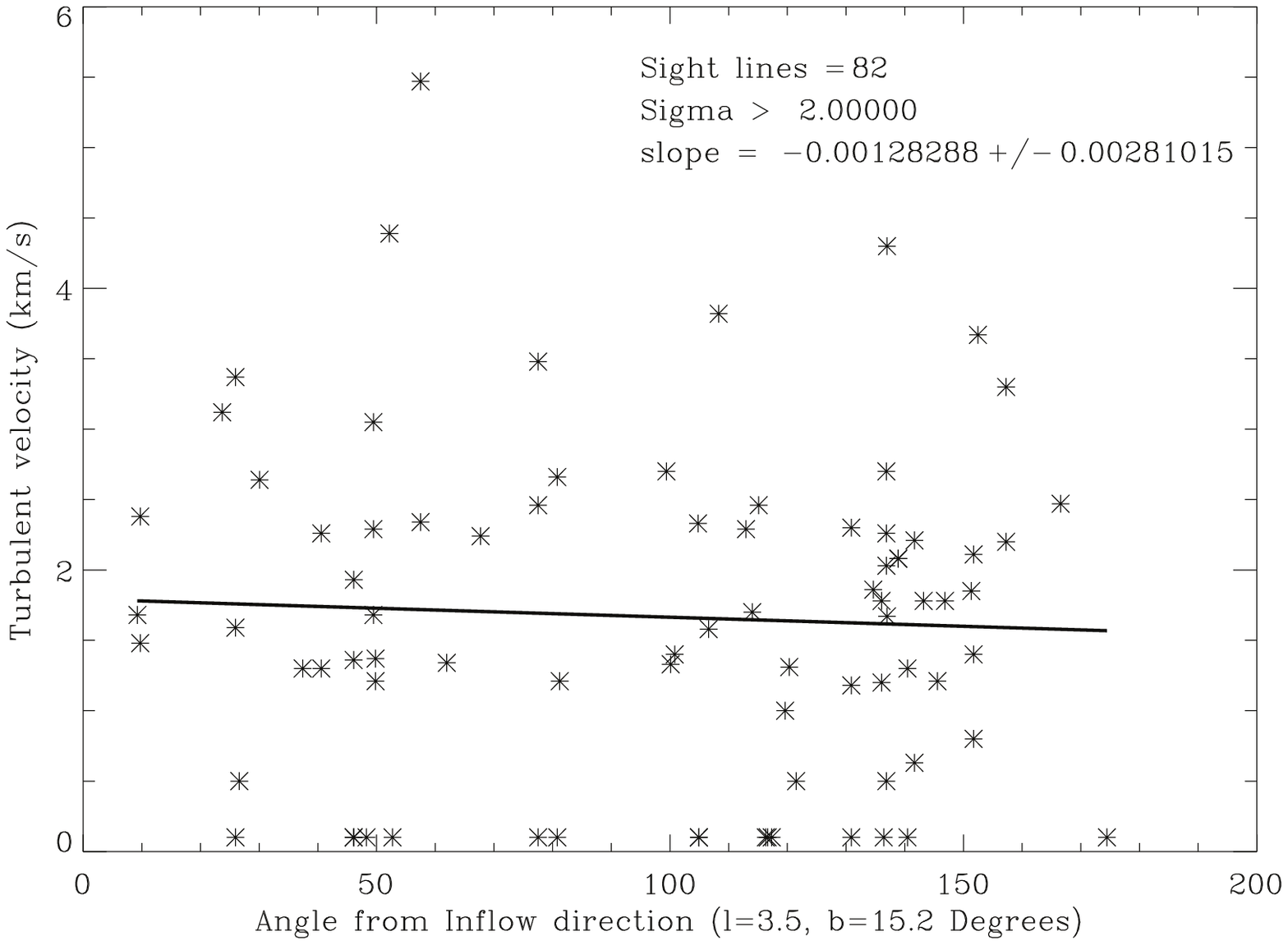}
\caption{{\bf Top}: Plot of sight line temperatures vs. angle from the LISM inflow direction. {\bf Bottom}: Plot of turbulent velocities vs. angle from the LISM inflow direction. Solid lines are least-squares linear fits to the data. The slope of the fit is 2.6 times its error indicating that the decrease with angle could be real.}
\label{Tvsinflow}
\end{figure}

A second test is whether temperatures and turbulent velocities are in any way related to the direction of inflowing interstellar gas. Figure~\ref{Tvsinflow} shows that temperatures and turbulent velocities do not depend on angle from the inflow direction.

\begin{figure}[h]
\centering
\includegraphics[width=10.0cm]{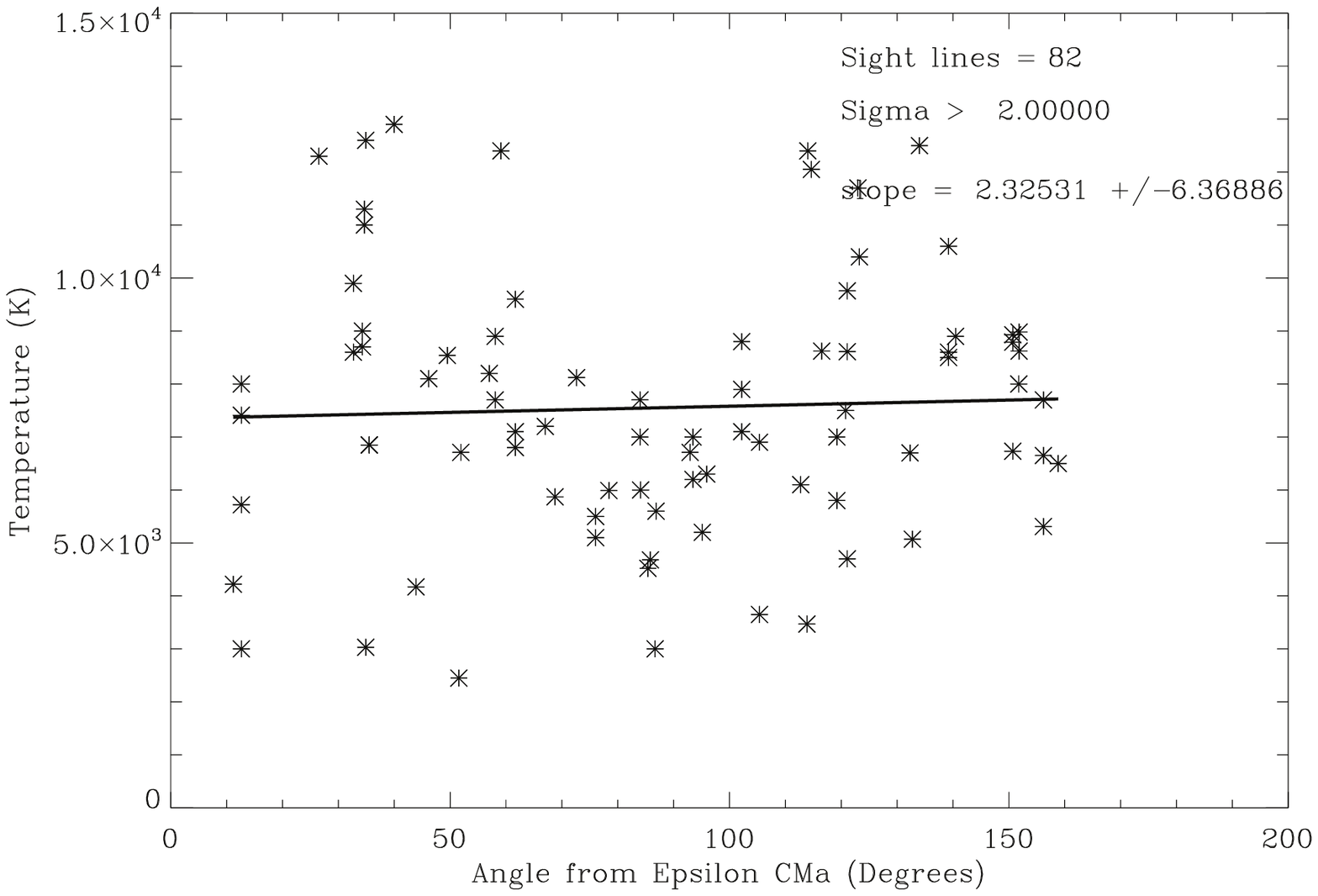}
\includegraphics[width=9.6cm]{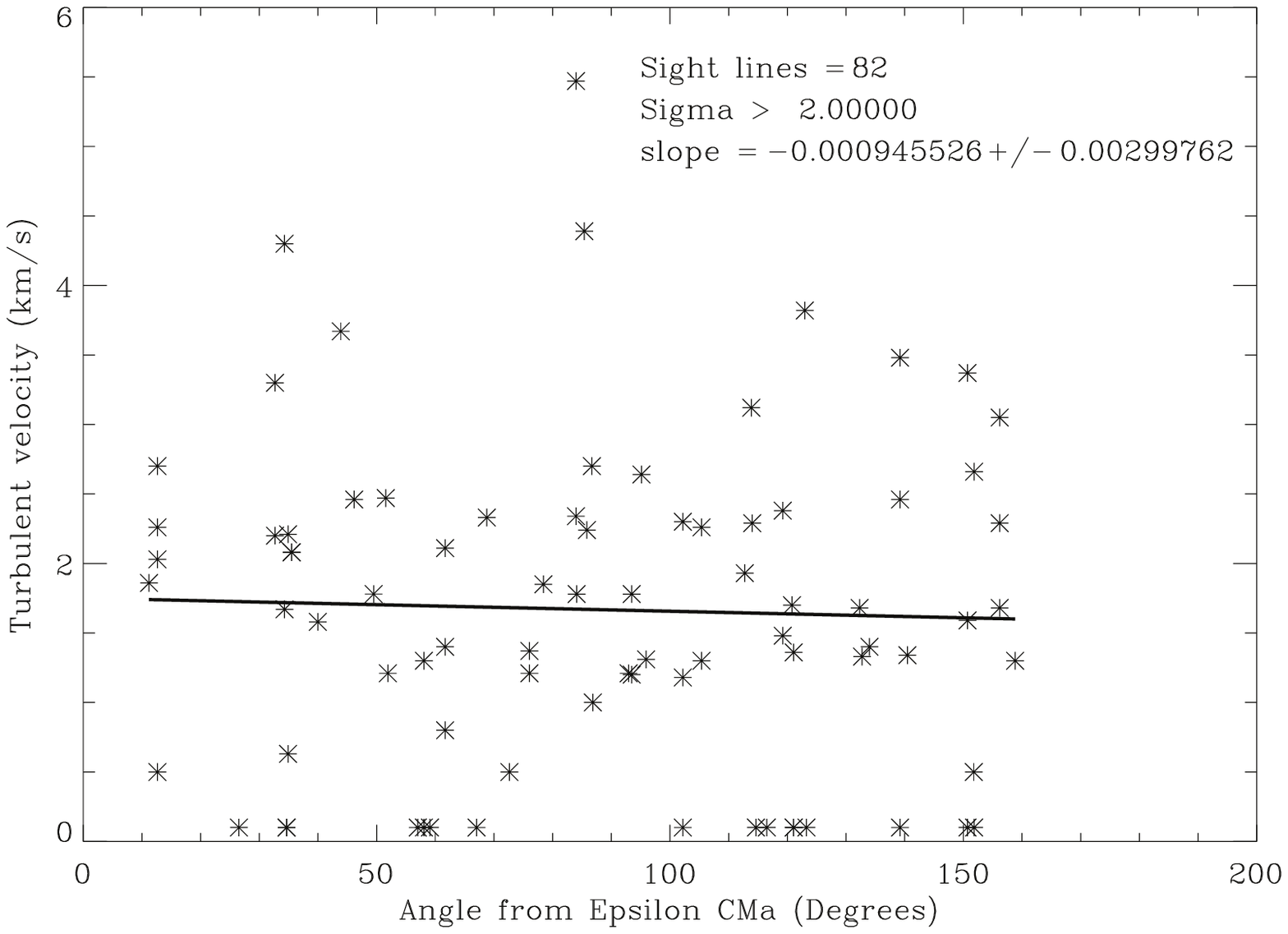}
\caption{{\bf Top}: Plot of sight line temperatures vs. angle from the star $\epsilon$~CMa.  {\bf Bottom}: Plot of sight line turbulent velocities vs. angle from $\epsilon$~CMa. Solid lines are least-squares linear fits to the data.}
\label{TvsECMa}
\end{figure}

We next consider whether the photoionizing radiation from the strongest extreme-UV (EUV) source in the sky, the star $\epsilon$~CMa is important. Figure~\ref{TvsECMa} shows no trend of temperatures or turbulent velocities with angle relative to $\epsilon$~CMa. The future development of a three dimensional model of the LISM would provide insight into where the shielding of EUV irradiation is important.

\begin{figure}[h]
\centering
\includegraphics[width=10.0cm]{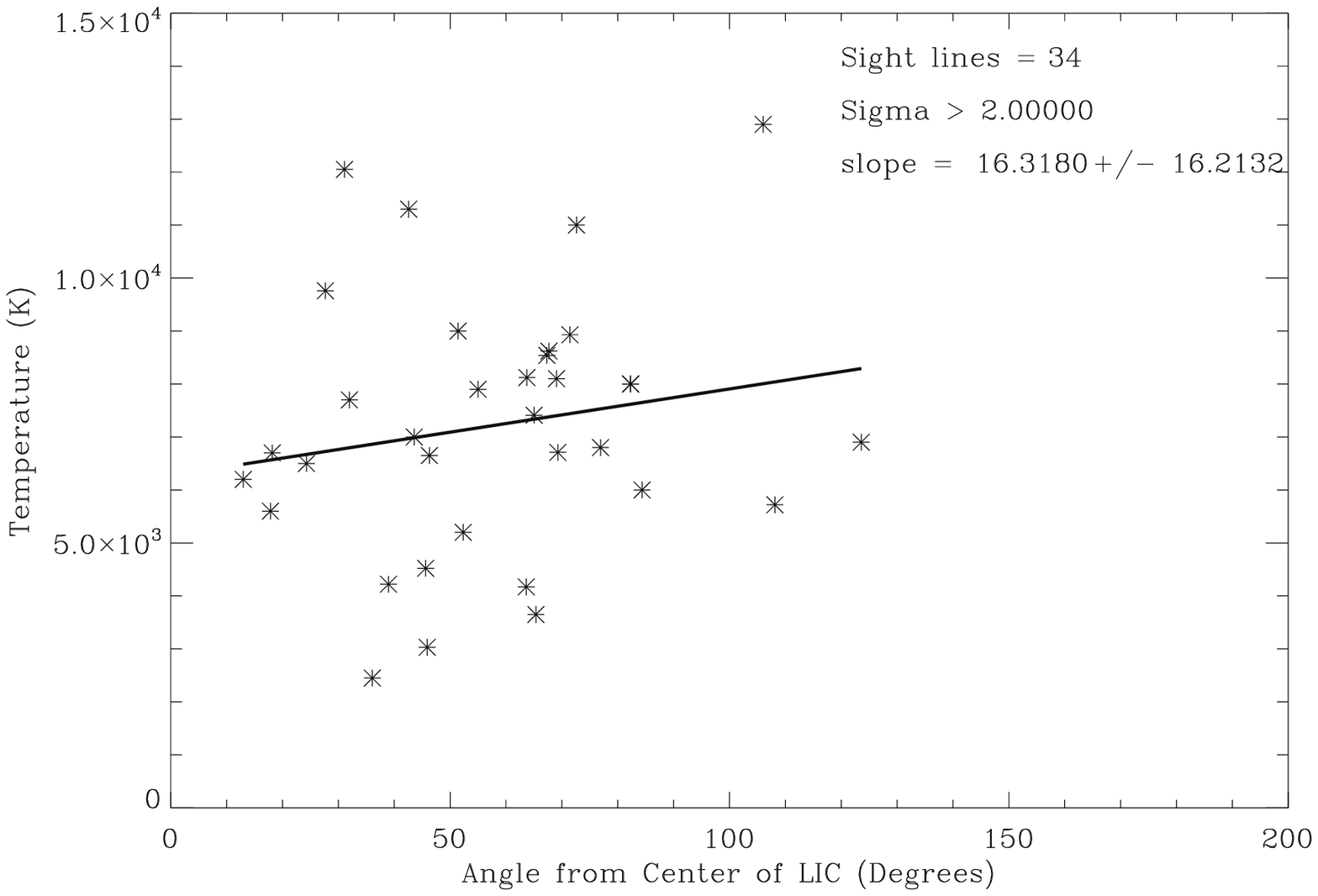}
\includegraphics[width=10.0cm]{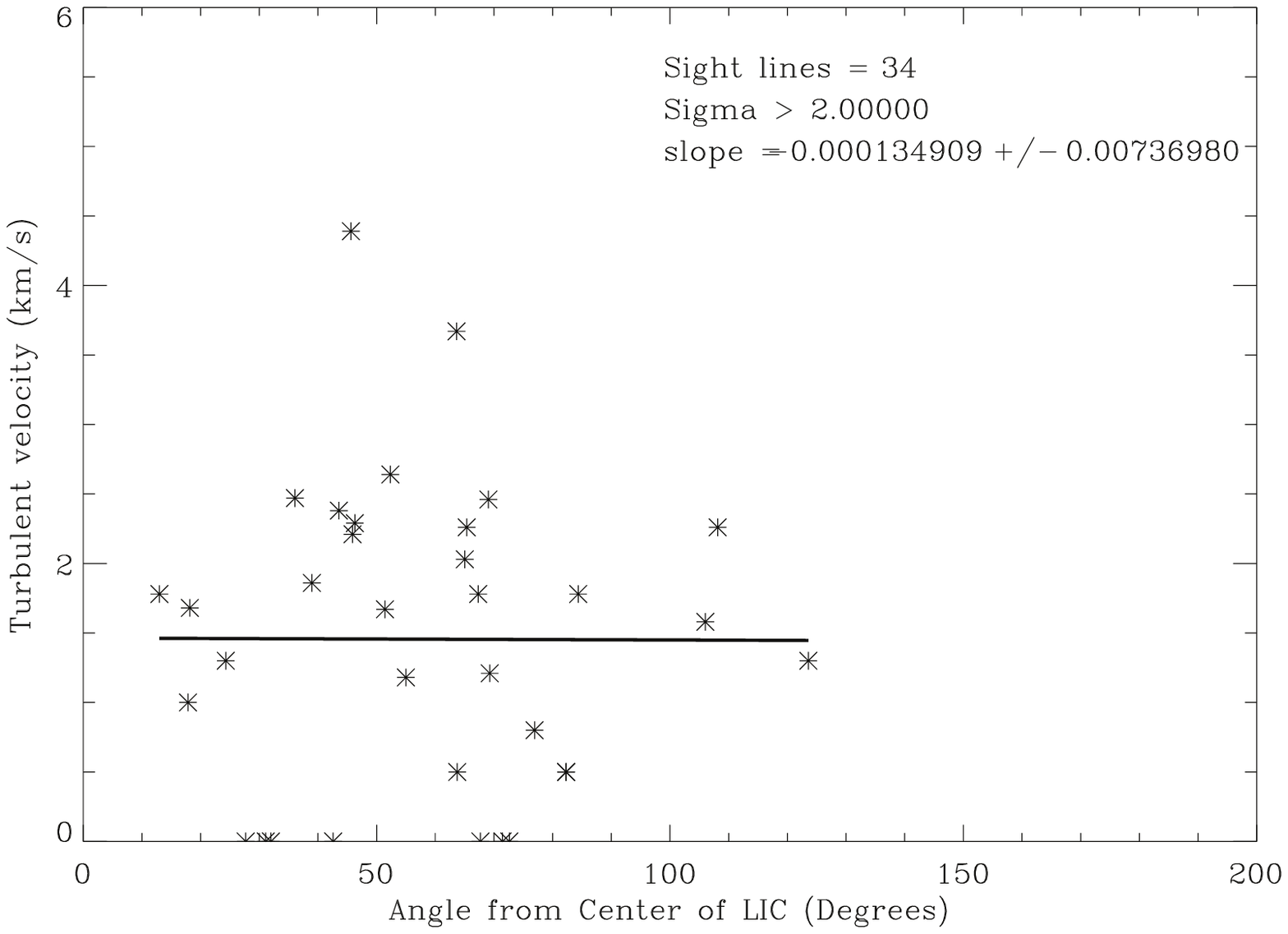}
\caption{{\bf Top}: Plot of sight line temperatures vs. angle from the center of the LIC. {\bf  Bottom}: Plot of sight line turbulent velocities vs. angle from the center of the LIC. Solid lines are least-squares linear fits to the data.}
\label{TvsLICangle}
\end{figure}

Finally, we consider whether temperatures and turbulent velocities are different in the core of the LIC compared to its edge. We test this by plotting in Figure~\ref{TvsLICangle} temperatures and turbulent velocities as a function of angle relative to the LIC core direction (l=$145^{\circ}$, b=$0^{\circ}$). There are no significant trends for either quantity implying that the LIC core and edge have no significant differences even though the range in these properties is very large.

\subsection{What is the Length Scale for Interstellar Spatial Variations?}

\begin{figure}[h]
\centering
\includegraphics[width=10.0cm]{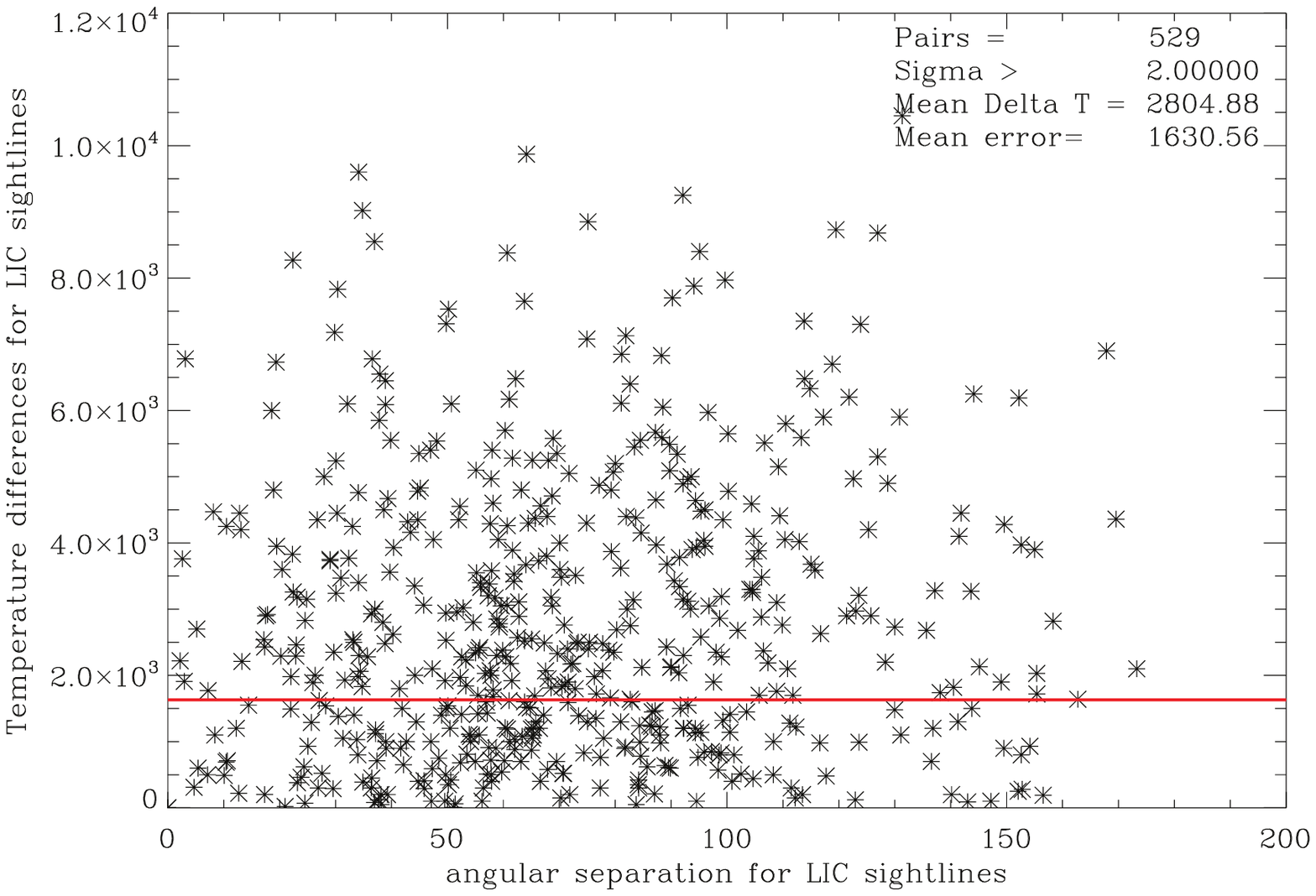}
\includegraphics[width=10.0cm]{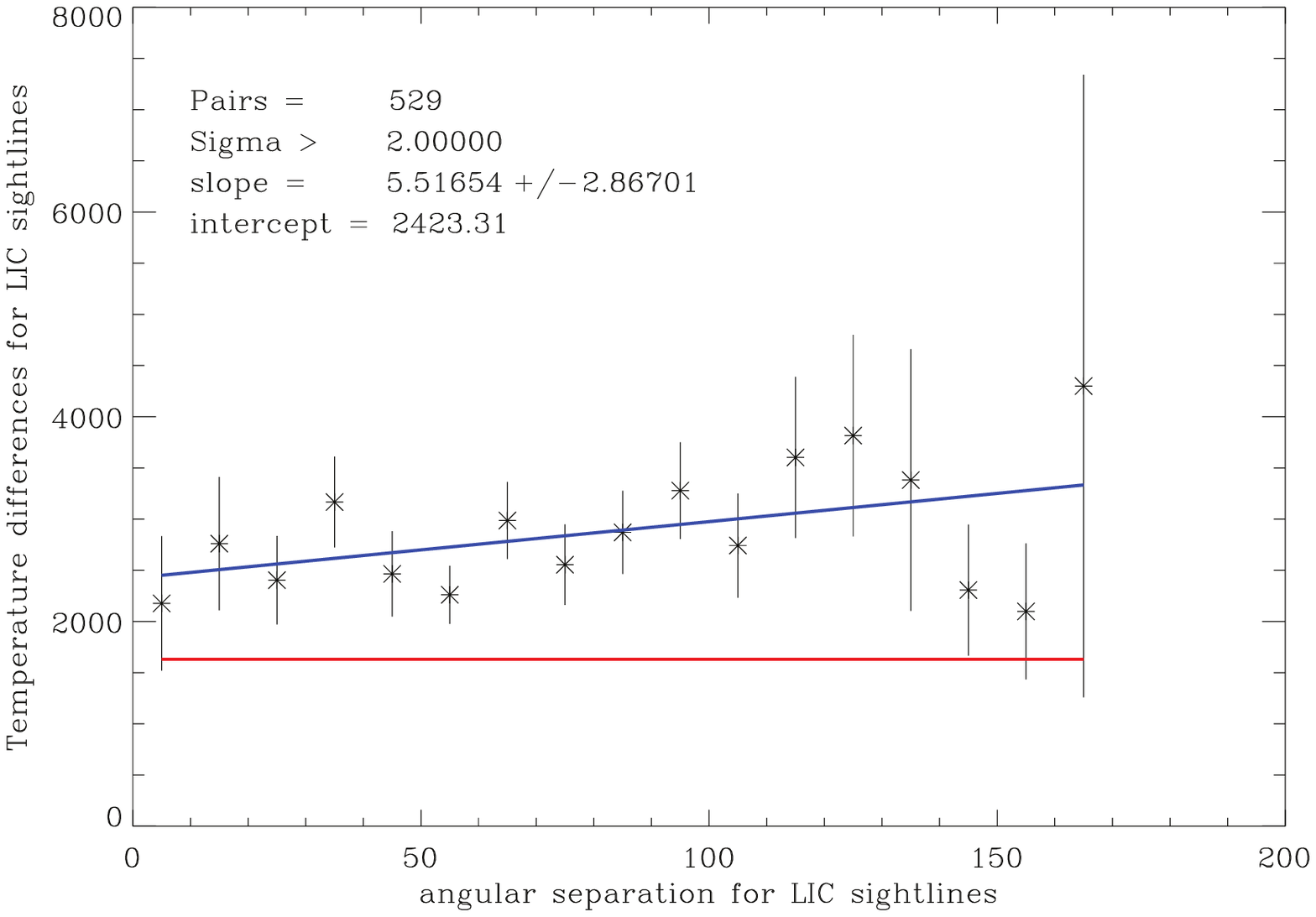}
\caption{{\bf Top}: Temperature differences for sight line pairs through the LIC as a function of angular separation between the sight lines. Only sight lines with temperatures at least two times their measurement errors are included. The horizontal red line is the mean uncertainty of individual temperature measurements. {\bf Bottom}: Mean temperature differences for sight line pairs through the LIC in $10^{\circ}$ angular bins. The vertical lines show the variance of the temperature differences in each angular bin and the star symbols are the mean values in each bin. The horizontal red line is the mean uncertainty of temperature measurements, and the blue line is a least squares fit to the mean values in each bin.}\label{Tsep}
\end{figure}

Another way of studying inhomogeneities inside the LIC is by comparing the temperature differences for all combinations of sight lines within the LIC. To minimize the role of measurement errors and possible systematic errors, Linsky et al. (2022) included only sight lines for which the temperature measurements are at least 2 times larger than the measurement errors. After subtraction of sight line pairs between two stars in a binary system, there are 529 unique sight line pairs. Figure~\ref{Tsep} shows the temperature differences for these pairs as a function of angular separation between the two sight lines. The temperature differences are as large as 9600~K, and most lie well above the mean measurement uncertainty of 1631~K. Thus for a large number of the sight line pairs, the temperature differences far exceed measurement uncertainties. 

The right side of the figure shows the temperature differences binned in $10^{\circ}$ wide angles between the two sight lines. The temperature differences increase with angular separation and are all above the measurement errors. The smallest angular scale of $2^{\circ}$.2 is for the Procyon-YZ CMi sight line pair, which has a temperature difference of 2718~K while the measurement uncertainty for this pair is only 415~K.  The halfway distance through the LIC towards Procyon is 0.5~pc, which implies a length scale for the separation of the sight lines in the LIC of 0.5tan(2$^{\circ}$.2)~pc =  4,000~AU. This length scale may be an upper limit. Since the Sun moves through the LIC at 5.1 AU/yr, the heliosphere could see changes in local interstellar properties within 600 years. 
We conclude that the angular scale for significant temperature differences  is at least as small as 4,000~au. 

\section{Pressure Balance between the Outer Heliosphere and Local Interstellar Cloud}

In this section we provide an overview of the pressure balance issues that will be presented in more detail in Paper~B. The size of the heliosphere and its three dimensional shape are controlled by the total pressure balance between the VLISM and LIC and then between the VLISM and heliosheath at the heliopause. The ionization of the inflowing interstellar gas also plays an important role in these balances. The heliosphere is now embedded in low density interstellar gas with $n$(H~I)$\approx 0.2$~cm$^{-3}$, but it has traversed both intercloud and supernova remnant regions with fully ionized hydrogen. The heliosphere may also have traversed high-density cold clouds such as the Local Leo Cold Cloud (Peek et al. 2011) with densities in excess of $10^4$~cm$^{-3}$ and pressures orders of magnitude larger than at present, but no direct evidence for this has been identified. In the latter case, the size of the VLISM would have shrunk to theorbits of Jupiter or even the Earth (Zank \& Frisch 1999, M\"uller et al. 2006). 

The estimate of the pressure balance that is relevant for the heliosphere involves two steps: (1) At the heliopause, pressure balance is established between the total pressure in the heliosheath and that in the VLISM. (2) The total pressure in the LIC (undisturbed by the presence of the heliosphere), including its ram pressure due to the motion relative to the Sun, ultimately determines the VLISM pressure where the flow is stopped at the heliopause. 

We start with the discussion of (1), limiting the evaluation to the location where the interstellar plasma flow stagnates in the upwind direction due to the combined action of the plasmas and the magnetic fields. The maximum pressure identified with {\em IBEX} ENAs (McComas and Schwadron 2014) likely occurs in this stagnation region. At any other location a more complex combination of pressures and flows would have to be considered. 

When looking at (2), we will consider which pressure components translate into the total pressure at the heliopause.
The total pressure consists of several components: cosmic-ray pressure $P$(cr), magnetic pressure $P$(mag), turbulent pressure $P$(turb), thermal pressure $P$(th), the pressure of supra-thermal ions $P$(supra-th), the pressure of hot (pick-up) ions $P$(hot-ions), and ram pressure $P$(ram). Not all of these components contribute significantly at all three locations. 
Pressure has units of dynes~cm$^{-2}$, but it is convenient to divide the pressure by Boltzmann's constant $k=1.38\times 10^{-16}$ erg~deg$^{-1}$, in which case $P/k$ has units of Kcm$^{-3}$ and is proportional to temperature times density. $P/k$ typically has values of a few thousand Kcm$^{-3}$ rather than $10^{-12}$ dynes~cm$^{-2}$.  

\subsection{VLISM Pressure at the heliopause}

The VLISM pressure components at the heliopause are compiled in Table~1. Some of these pressures have been directly measured by the {\em Voyager} spacecraft but others come from models. Beyond the heliopause, {\em Voyager-1} measured Galactic cosmic rays above 3 MeV per nucleon with a broad maximum in the energy spectrum at 10--50 MeV per nucleon (Cummings et al. 2016). The energy density $E/V$=0.83--1.02 eV cm$^{-3}$ corresponds to a non-relativistic pressure $P$(cr)/$k=2E/3Vk=7,150\pm730$ Kcm$^{-3}$. This value of $P$(cr)/$k$ was measured where the effects of the solar magnetic field are small. Since {\em Voyager~1} and {\em Voyager~2} detected no radial gradient in the cosmic ray pressure (Stone 2019), we assume that $P$(cr) has the same value in the VLISM and LIC. For simplicity, we also assume that the sum of the pressures of galactic and of anomalous cosmic rays in the heliosheath equal $P$(cr)/$k$ outside the heliopause. 

Measurements of the magnetic field by {\it Voyager 1} after passage through the heliopause show values of $B = 4.6\pm 0.3 \mu$G, obtained from the figures in Burlaga et al. (2021) at the heliopause and a very slow linear decrease in the magnetic field strength with distance reaching 
$B=4.0 \mu$G at the end of 2020 (149.2~au from the Sun). This result is an extension and re-calibration 
(Burlaga, private communication) of the results published by Burlaga et al. (2021). The magnetic pressure at the heliopause is $P$(mag)=$B^2/8\pi k = 6100\pm 820$~Kcm$^{-3}$.
At the stagnation point, the thermal plasma pressure $P$(th) almost completely consists of thermal pressure by H$^+$, He$^+$ and electrons. Because no direct measurements of this pressure are available, we adopt values found for the heliopause in model 2 of Zank et al. (2013) ($T = 28,000$~K, $n$(H$^+$) = 0.09~cm$^{-3}$ + 10\% of He$^+$ and $n$(e) matching for charge neutrality) as the values chosen for the LIC parameters are closest to most recent results from observations. The fluxes of suprathermal particles on the {\em Voyagers} has dropped in excess of two orders of magnitude beyond the heliopause, likely making this contribution insignificant. In the models computed by Zank et al. (2013) the pressure at 300~au in the supra-thermal tail of the velocity distribution is small, $P$(supra-th)/$k=160\pm 20$~Kcm$^{-3}$ (Haoming Liang private comm.). Dialynas et al. (2021) measured the 40-139 keV flux of pickup ions just outside the the heliopause from {\em Voyager-1} data, but the pressure is very small, $P$(hot-ions)/$k=2.17\pm 0.31 $~Kcm$^{-3}$ (Dialynas private comm.).

\subsection{Pressure in the heliosheath}

The pressure components in the heliosheath are compiled in Table~2, starting with the cosmic ray pressure mentioned above. Schwadron et al. (2011) estimated the total pressure in the heliosheath from maps of the energetic neutral atom (ENA) flux obtained from the {\em Interstellar Boundary Explorer (IBEX)} satellite located near 1~au. The ENA flux measurements refer to the globally distributed flux after subtracting the ribbon flux and the loss of ENAs between the heliosheath and 1~au from charge exchange reactions with solar wind protons. The inferred total plasma pressure is $P$(plasma) = 1.9~pdynes cm$^{-2}$  or $P$(plasma)/$k$= 13,800 Kcm$^{-3}$ near the nose direction, assuming a heliosheath thickness of 38~au. 
The plasma pressure, $P$(plasma)/$k$ includes both thermal and non-thermal ($P$(supra-th) and $P$(hot-ion)) components and the dynamic pressure of the outflowing plasma downstream of the heliopause. The plasma pressure matches almost exactly the sum of thermal heliosheath pressure up to keV (13,200 Kcm$^{-3}$), as obtained from {\em IBEX} ENA flux observations at the pressure maximum (McComas and Schwadron 2014), adjusted for a heliosheath thickness of 38~au, and the suprathermal and hot ion pressures from INCA ENA and {\em Voyager~1} ion observations (Dialynas et al. 2021).

\subsection{Pressure in the undisturbed LIC}

The pressure components in the undisturbed LIC are compiled in Table~3, starting again with the cosmic ray pressure from above. Analysis of the {\em IBEX} ribbon data by Zirnstein et al. (2016) resulted in a best fit magnetic
field strength  $B=2.93\pm 0.07 \mu$G corresponding to $P$(mag)/$k=2,470\pm 120$~Kcm$^{-3}$. We take this value as an estimate of the magnetic field strength in the LIC. The thermal pressure, $P$(th)/$k=n$(tot)$T=2,740\pm750$~Kcm$^{-3}$, is computed from the mean weighted temperature $T = 7,563\pm 2,077$~K in the LIC sight lines (Figure 5). The turbulent pressure, $P$(turb)/$k=\rho v^2$(turb)=$280^{+290}_{-190}$~Kcm$^{-3}$, is computed from $v$(turb) = $2.13\pm 0.93$~km~s$^{-1}$ (Figure 5) and $n$(tot)=$0.362\pm 0.02$~cm$^{-3}$ using the densities in Model 26 of Slavin \& Frisch (2008). Note that in both the turbulent and ram components a 10\% density contribution from He translates into a 40\% pressure contribution in addition to the H component pressure and thus is not insignificant. Summing up these component pressures, we arrive at a total LIC pressure in its rest frame of $P$(total)=$12,640\pm 1,050$~Kcm$^{-3}$, a value much lower than the estimated total pressures inside (heliosheath) and outside (VLISM) of the heliopause. This deficiency points at the importance of the ram pressure $P$(ram)$=\rho v^2$ of the LIC relative to the VLISM for the overall heliospheric pressure balance.
 
The ram pressure of the LIC that impacts the VLISM is the forward momentum per unit area lost by particles from the LIC that interact with plasma in the VLISM rather than passing through unimpeded or being deflected around the heliopause with little forward momentum loss. The inflows of all interstellar charged particles (electrons, protons and ions) are diverted around the heliopause by the magnetic field, and thus transfer forward momentum to the VLISM only via the magnetic field. A portion of the forward momentum  adds to the thermal plasma pressure via density compression and heating, while the rest ends up in the diverted flow around the heliopause. A fraction of the inflowing hydrogen atoms charge exchange with the diverted and slowed down interstellar plasma flow adding to the momentum balance in this plasma, while a fraction penetrate through the VLISM without interactions and thus do not contribute to the ram pressure. Since helium atoms pass through the VLISM with few interactions, their contribution to the ram pressure in the VLISM is also minimal. 

If all of the inflowing LIC particles were to contribute to the ram pressure on the heliosphere produced by the $25.9\pm 0.2$ km~s$^{-1}$ speed of the Sun through the LIC measured by neutral helium atoms flowing into the heliosphere (e.g., Swaczyna et al., 2022, Wood et al. 2015), then $P$(ram)/$k = 24, 700 \pm 2, 400$~Kcm$^{-3}$. The total LIC pressure projected at the heliosphere, including this ram pressure, would then amount to $P$(total)/$k = 37, 340 \pm 2, 620$~Kcm$^{-3}$, substantially in excess of the pressures at the heliopause. To estimate the fraction of the original LIC ram pressure that contributes to the pressure on the heliopause would require extensive modeling. This fraction is implicit in the combined magnetic pressure, obtained from Voyager measurements, and thermal plasma pressure, obtained from the model by Zank et al. (2013) listed in Table~1, $P$(mag)/$k$ + $P$(th)$k= 11,680\pm 1,500$~ Kcm$^{-3}$. When compared with the combined magnetic and thermal pressures listed in Table~3, $P$(mag)/$k$ + $P$(th)$k= 5,210 \pm 760$~Kcm$^{-3}$, we obtain as estimate for the effective portion of the LIC ram pressure as the difference between these the two values, $P$(rameff)/$k= 6,470\pm 1,680$~Kcm$^{-3}$. The total pressure in the LIC is then $P$(total)/$k=19,110\pm 1,980$~Kcm$^{-3}$.

For comparison we mention the different approach employed by Schwadron et al. (2011) to estimate the ram pressure needed to establish pressure balance between the heliosheath and the LIC as 0.79 pdynes cm$^{-3}$, corresponding to $P$(ram)/$k=5,730$~Kcm$^{-3}$ and $P$(total)/$k=18,366$~Kcm$^{-3}$. Because both of these two estimates are not based on an accurate bottoms-up calculation, more detailed work is required in the future.
 
 \subsection{Pressure comparisons}

We intercompare the total pressures in the heliosheath, VLISM just outside of the heliopause, and the LIC to determine whether there may be significant imbalances that would cause relative flows. The total pressure outside the heliopause ($18,830\pm 1,670$ Kcm$^{-3}$) appears to fall short by about 10\% of the pressure inside ($20,400\pm 780$~Kcm$^{-3}$), although the uncertainties in these pressures and possible other pressure sources that are not included are consistent with total pressure balance and therefore no flows. With the inclusion of ram pressure, the total pressure in the LIC as seen from the VLISM ($19,110\pm 1,980$~Kcm$^{-3}$) is consistent with the total pressure in the VLISM within the uncertainties.

We next compare the total pressures in the heliosheath, VLISM outside of the heliopause, and the LIC with the Galactic neighborhood. Cox (2005) estimated the weight of overlying material perpendicular to the Galactic plane. Since the Sun is very close to the Galactic plane, the gravitational pressure is $P$(grav)=$3.0\times 10^{-12}$ dynes~cm$^{-2}$ or $P$(grav)/$k=22,000$~Kcm$^{-3}$. The total pressure in the heliosheath is close to $P$(grav)/$k$, and the total pressure in the VLISM may be slightly below. The total internal pressure in the LIC, $12,636\pm 1,046$, however, is far below $P$(grav)/$k$, but the LIC is likely moving with respect to its external environment producing a ram pressure that must be added to its total  internal pressure. This question will be addressed in Paper B, but a very rough estimate of $P$(ram) follows from assuming that the LIC is moving with respect to the local standard of rest, $v$(LSR)=$17.9\pm 3.27$~km~s$^{-1}$ (Frisch et al. 2011). In this case, $P$(ram)(LIC)/$k=19,000^{+8,000}_{-6,480}$~Kcm$^{-3}$. For comparison, Snowden et al. (2014) estimated that the thermal pressure in the Local Bubble is 10,700~Kcm$^{-3}$. Linsky \& Moebius (2022) will consider the pressure balance with Local Bubble in detail.

\begin{table}[ht]
\begin{center}
\begin{minipage}{12cm}
\caption{Pressures components at the Heliopause(Kcm$^{-3}$)}\label{VLISMPressure}%
\begin{tabular}{@{}lll@{}}
\toprule
Component & Parameter & Component Pressure\\
\midrule
$P$(cr)/$k$ & 0.83--1.02 eV/cm$^3$ & $7,150\pm730$\\
$P$(mag)/$k$ & $4.6\pm 0.3\mu$G & $6,100\pm 820$\\
$P$(plasma)/$k$ & model with $T= 28,000$~K & $5,580\pm 1,260$\\
$P$(total)(HP)/$k$ &  & $18,830\pm 1,670$\\
\botrule
\end{tabular}
\end{minipage}
\end{center}
\end{table}
\bigskip

\begin{table}[ht]
\begin{center}
\begin{minipage}{12cm}
\caption{Pressures components in the Heliosheath (Kcm$^{-3}$)}\label{HSPressure}%
\begin{tabular}{@{}lll@{}}
\toprule
Component & Parameter & Component Pressure\\
\midrule
$P$(cr)/$k$ & 0.83--1.02 eV/cm$^3$ & $7,150\pm730$\\
$P$(plasma)({\em IBEX})/$k$ & 0.5-6 keV ENA fluxes & $5,180\pm260$\\
$P$(plasma)({\em INCA + Voyager})/$k$ & 10--150 keV ions & 7,970\\
 $P$(total)(HS)/$k$ & & $20,400\pm 780$\\
\botrule
\end{tabular}
\end{minipage}
\end{center}
\end{table}
\bigskip

\begin{table}[h]
\begin{center}
\begin{minipage}{12cm}
\caption{Pressure Components in the LIC (Kcm$^{-3}$)}\label{LICPressure}%
\begin{tabular}{@{}lll@{}}
\toprule
Component & Parameter & Component Pressure\\
\midrule
$P$(cr)/$k$ & 0.83--1.02 eV/cm$^3$ & $7,150\pm730$\\
$P$(mag)/$k$ & $2.93\pm 0.07 \mu$G & $2,470\pm 120$\\
$P$(turb)/$k$ & $v$(turb)=$2.13\pm0.93$ km/s & $280^{+290}_{-190}$\\
$P$(th)/$k$ & $T=7,563\pm 2,077$ K & $2,740\pm750$\\
$P$(total)(at rest) & & $12,640\pm 1,050$\\ 
\midrule
$P$(rameff)(VLISM)/$k$ &  &  $6,470\pm 1,680$\\
$P$(total)(VLISM)/$k$ & Kcm$^{-3}$ & $19,110\pm 1,980$\\
\botrule
\end{tabular}
\end{minipage}
\end{center}
\end{table}
\bigskip

\section{What would happen to the heliosphere if/when it enters the intercloud medium?}

The inter-cloud medium, if it exists, would have to contain fully ionized hydrogen in order to be undetected by H~I Lyman-$\alpha$ absorption. Since the inflow of neutral hydrogen into the heliosphere is essential for charge exchange reactions, a future heliosphere  embedded in an inter-cloud medium would not have a hydrogen wall and would have very different ionization. It is likely that hot ions would not be present.

M\"uller et al (2006) computed heliospheric models for a wide range of interstellar environments including the case of a hot Local Bubble environment with fully ionized hydrogen. For this case there are no charge exchange reactions with the inflowing plasma and, therefore, no pickup ions to heat the solar wind, no hydrogen wall, and a very wide heliosheath of million degree plasma extending from a termination shock at 90~au to a heliopause near 300~au.  The absence of heating by pickup ion reactions results in adiabatic expansion and resulting cooling of the solar wind inside of the termination shock and then a steep jump in temperature into the million degree heliosheath. If, instead, the inter-cloud medium is similar to the warm ($T$=10,000-20,000~K) photo-ionized Local Cavity as described by Linsky \& Redfield (2021), then the heliosphere would likely be similar to the hot environment model, except that the plasma in the heliosheath and beyond the heliopause would be relatively cool in the absence of charge exchange reactions and hot inflowing plasma. For a future heliosphere embedded in a hot inter-cloud medium, $P$(th) will be much larger than for the LIC, but $P$(ram) would be smaller because the completely ionized hydrogen would be deflected around the distant heliopause departing little momentum to the heliosphere. It is not obvious which term would dominate the external total pressure to determine the size of the future heliosphere. If a future heliosphere is embedded in a warm inter-cloud medium, then both $P$(th) and $P$(ram) will be smaller than at present and the heliosphere will be larger than its present value.
Realistic heliospheric models are needed to test this prediction.

\section{Conclusions}

(1) The kinematical model of the LISM predicts the past and future trajectory of the heliosphere through partially ionized clouds and the postulated inter-cloud medium with fully ionized hydrogen. The properties of the heliosphere will be very different depending upon whether the inflowing interstellar gas contains neutral hydrogen or fully ionized hydrogen. The kinematical model computes cloud sizes from interstellar hydrogen column densities and the assumption that the neutral hydrogen number density is  
$n$(H~I)=0.20~cm$^{-3}$, the same as the LIC close to the heliosphere. This assumption may not be valid for other clouds or even most of the LIC. If instead $n$(H~I)$\approx 0.10$~cm$^{-3}$ is a typical density in clouds, the clouds will be larger and completely fill the available space with no room for an inter-cloud medium. If/when a future heliosphere moves into an inter-cloud environment, the absence of inflowing neutral hydrogen will result in a cooler solar wind and fewer energetic particles in the heliosheath and beyond. The size of the future heliosphere will also depend on the temperature of the surrounding medium. Thus the future evolution of the heliosphere depends critically upon only a factor of 2 in the unknown neutral hydrogen density of the G cloud that the heliosphere could enter in less than 2000 years.  

(2) The analysis of an increasing number of interstellar velocity components and their physical properties shows that there is a large range of temperatures and turbulent velocities within the LIC and other clouds. The temperature and turbulent velocity inhomogeneities appear to be random and their distributions can be fit by Gaussians. Temperatures and turbulent velocities may not be spatially correlated in the same way as the kinematics of LISM clouds, but instead could be controlled by the ionization source direction and self-shielding. There are no signifiant trends with distance to the background star or angle from the inflow direction. There are no correlations with angle from the main photo-ionizing source 
$\epsilon$~CMa, and no differences between the LIC core and its outer regions. Also, what previous analyses have called temperature may in reality be a combination of thermal and supra-thermal velocities that broadens absorption lines produced by  low mass atoms (e.g., D I) but not high mass ions (e.g., Mg~II and Fe~II).

(3) The temperature differences between LIC sight lines increase with separation and are much larger than the measurement errors. The smallest angular separation of 2$^{\circ}.2$ has a temperature difference of 2718~K compared to a measurement error of 415~K. This sets an angular scale for the temperature inhomogeneities that corresponds to a length scale smaller than 4,000~au in the LIC. Since the Sun moves through the LISM at 5.1 au/yr, the heliosphere could see changes in the surrounding interstellar gas within 600 years. 

(4) We tabulate the measured or estimated pressures due to cosmic rays, magnetic fields, turbulence, temperature, supra-thermal ions, and hot (pickup) ions in the heliosheath, the VLISM just outside of the heliopause, and the LIC. For the heliosheath and VLISM, the total pressures are approximately equal to the gravitational pressure of overlying material in the Galaxy. By including the effective ram pressure manifest in the increased temperature and magnetic field strength at the heliopause, the total pressure in the VLISM appears to be consistent with that in the heliosheath and with the gravitational pressure, but extensive modeling is needed. The total internal pressure in the LIC is far below that of the VLISM, but the LIC is likely moving with respect to its environment producing a ram pressure term that must be added to its internal pressure to test for pressure balance with its environment.

\bmhead{Acknowledgments}

This work was made possible by the International Space Science Institute and its interdisciplinary workshop "The Heliosphere in the Local Interstellar Medium".
We acknowledge support from the NASA Outer Heliosphere Guest Investigators Program to Wesleyan University and the University of Colorado through grant 80NSSC20K0785. We thank Haoming Liang and G. Zank for measuring the supra-thermal pressure in the VLISM from a figure in the Zank et al. (2013) paper, and K. Duilynas for input on the hot-ion pressure. We also thank A. Cummings for his comments, and L. Burlaga for a personal communication concerning magnetic fields in the heliosphere. 

Facilities: {\em HST}(STIS), {\em HST}(HRS),  {\em EUVE}, {\em Voyager I, Voyager II}

 \end{document}